\documentclass[useAMS,usenatbib]{mn2e}

\usepackage{times}
\usepackage{graphicx, latexsym, amssymb, amscd, psfrag}
\usepackage{epsfig}
\usepackage{amsmath}
\usepackage{lineno}

\title[H-ATLAS: The environment of low red shift galaxies]
{The environment and characteristics of low redshift galaxies 
detected by the {\it Herschel}-ATLAS}
\author[A. Dariush]{A. Dariush$^{1,20}$\thanks{E-mail: ali.dariush@astro.cf.ac.uk (AAD)} , 
L. Cortese$^{13}$, S. Eales$^{1}$, E. Pascale$^{1}$, M. W. L. Smith$^{1}$, L. Dunne$^{2}$,  	   
\newauthor
S. Dye$^{1,2}$, D. Scott$^{30}$, R. Auld$^{1}$, M. Baes$^{6}$, J. Bland-Hawthorn$^{31}$,S. Buttiglione$^{9}$,  A. Cava$^{3}$,      
\newauthor 
 D.L. Clements$^{20}$, A. Cooray$^{21}$,G. DeZotti$^{9,26}$, S. Driver$^{11}$, J. Fritz$^{6}$, H.L. Gomez$^{1}$,   A. Hopkins$^{12}$,  
\newauthor
  R. Hopwood$^{20}$,R.J. Ivison$^{7,14}$, M.J. Jarvis$^{22,29}$, D.H. Jones$^{12}$, L. Kelvin$^{11}$, H. G. Khosroshahi$^{32}$,  
\newauthor
   J. Liske$^{13}$, J. Loveday$^{15}$, S. Maddox$^{2}$, B.F. Madore$^{28}$, M.J. Micha{\l}owski$^{14}$, P. Norberg$^{14}$,    
\newauthor
 S. Phillipps$^{4}$, M. Pohlen$^{1}$, C.C. Popescu$^{18}$, A.S.G. Robotham$^{11}$, G. Rodighiero$^{27}$,  M. Prescott$^{10}$,  
\newauthor
 E. Rigby$^{2}$,  M. Seibert$^{28}$, D.J.B. Smith$^{2}$,   P. Temi$^{8}$, R.J. Tuffs$^{19}$, P. van der Werf$^{14,23}$\\ \ \\ 
$^{1}$School of Physics and Astronomy, Cardiff University, the Parade, Cardiff, CF24 3AA, UK\\
$^{2}$School of Physics and Astronomy, University of Nottingham, University Park, Nottingham NG7 2RD, UK\\
$^{3}$Departamento de Astrof\'{\i}sica, Facultad de CC. F\'{\i}sicas, Universidad Complutense de Madrid, E-28040 Madrid, Spain\\
$^{4}$Astrophysics Group, H.H. Wills Physics Laboratory, University of Bristol, Tyndall Avenue, Bristol BS8 1TL, UK\\
$^{5}$Laboratoire d'Astrophysique de Marseille, UMR6110 CNRS, 38 rue F. Joliot-Curie, F-13388 Marseille France\\
$^{6}$Sterrenkundig Observatorium, Universiteit Gent, Krijgslaan 281 S9, B-9000 Gent, Belgium\\
$^{7}$UK Astronomy Technology Centre, Royal Observatory, Edinburgh, EH9 3HJ, UK\\
$^{8}$Astrophysics Branch, NASA Ames Research Center, Mail Stop 245-6, Moffett Field, CA 94035, USA\\
$^{9}$INAF Osservatorio Astronomico di Padova, Vicolo Osservatorio 5, I-35122, Padova, Italy\\
$^{10}$Astrophysics Research Institute, Liverpool John Moores University, 12 Quays House, Egerton Wharf, Birkenhead, CH41 1LD, UK\\
$^{11}$SUPA, School of Physics and Astronomy, University of St. Andrews, North Haugh, St. Andrews, KY16 9SS, UK\\
$^{12}$Australian Astronomical Observatory, PO Box 296, Epping, NSW, 1710, Australia\\
$^{13}$European Southern Observatory, Karl-Schwarzschild-Strasse 2 D-85748, Garching bei M¨unchen, Germany\\
$^{14}$SUPA, Institute for Astronomy, University of Edinburgh, Royal Observatory, Blackford Hill, Edinburgh EH9 3HJ, UK\\
$^{15}$Astronomy Centre, Department of Physics and Astronomy, University of Sussex, Falmer, Brighton, BN1 9QH, UK\\
$^{18}$Jeremiah Horrocks Institute, University of Central Lancashire, Preston, PR1 2HE, UK\\
$^{19}$Max-Planck-Institut f\"ur Kernphysik, Saupfercheckweg 1, D-69117, Heidelberg, Germany\\
$^{20}$Physics Department, Imperial College London, Prince Consort Road, London, SW7 2AZ, UK\\
$^{21}$Department of Physics and Astronomy, UC Irvine, Irvine, CA 92697, USA\\
$^{22}$Centre for Astrophysics Research, Science \&\ Technology Research Institute, University of Hertfordshire, Hatfield, Herts, AL10 9AB, UK\\
$^{23}$Leiden Observatory, Leiden University, PO Box 9513, NL - 2300 RA Leiden, The Netherlands\\
$^{24}$International Centre for Radio Astronomy Research, The University of Western Australia, 35 Stirling Hwy, Crawley, WA 6009, Australia\\
$^{26}$SISSA, Via Bonomea 265, I-34136 Trieste, Italy\\
$^{27}$University of Padova, Vicolo Osservatorio 3, I-35122 Padova, Italy\\
$^{28}$Observatories of the Carnegie Institution of Washington, 813 Santa Barbara Str, Pasadena, CA91101, USA\\
$^{29}$ Physics Department, University of the Western Cape, Cape Town, 7535, South Africa\\
$^{30}$Department of Physics \& Astronomy, University of British Columbia, 6224 Agricultural Road, Vancouver, BC, V6T1Z1, Canada\\
$^{31}$Sydney Institute for Astronomy, University of Sydney, NSW 2006, Australia\\
$^{32}$IPM School of Astronomy, Larak Garden, opposite Araj, Artesh Highway,Tehran, Iran\\
}

\begin{document}



\maketitle

\label{firstpage}


\begin{abstract}
 
We investigate the ultraviolet and optical properties and environment
of low redshift galaxies detected in the {\it Herschel} Astrophysical
Terahertz Large Area Survey (H-ATLAS) science demonstration data. We
use the Sloan Digital Sky Survey seventh release and the Galaxy And
Mass Assembly database to select galaxies with
$r_{_{\rm Petro}} \leq 19.0$~mag in the redshift range $0.02 \leq z
\leq 0.2$ and look for their submillimeter counterparts in H-ATLAS.
Our results show that at low redshift, H-ATLAS detects mainly
blue/star-forming galaxies with a minor contribution from red systems
which are highly obscured by dust. In addition we find that
the colour of a galaxy rather than the local density of its
environment determines whether it is detectable by H-ATLAS. The
average dust temperature of galaxies that are simultaneously detected
by both PACS and SPIRE is $25\pm4$~K, independent of environment.
This analysis provides a glimpse of the potential of the H-ATLAS data to
investigate the submillimeter properties of galaxies in the local
universe.

\end{abstract}

\begin{keywords}

\end{keywords}


\section{Introduction}
\label{introduction}

The wide range of observed physical characteristics of galaxies is
indicative of a large variance in galaxy formation history.
Observations show that the environment of a galaxy plays an important role
in shaping its observed properties. \cite{dressler} showed that
galaxy morphology is a strong function of galaxy density and numerous
studies since then have demonstrated the dependence of galaxy
properties on local environment
\citep{lewis,gomez,balogha,baloghb,blanton,park,mill,ball,lee}.

Results from large sky surveys such as the Sloan Digital Sky Survey
(SDSS) have revealed that the distribution of galaxy colours is
bimodal, i.e., the so called `blue cloud' versus the `red sequence'
\citep{stra,baldry}. The relative numbers of blue and red galaxies at
a fixed luminosity are observed to vary strongly with local density,
with blue and red galaxies predominantly populating low and high
density environments, respectively ~\citep[e.g.][]{baloghb,ball}.  The
major factors that influence the observed colour of a galaxy are its
star formation history (SFH), the amount of dust attenuation, and
metallicity \citep{johnson06,johnson07}.  The vast majority of
galaxies in the blue cloud are actively forming stars while the red
sequence consists mainly of early-type passive galaxies (i.e., having
little or no current star formation) with additional minor
contributions from heavily obscured star-forming galaxies or edge-on
systems.

Analyses of the dust attenuation in star-forming galaxies suggest that
in comparison to quiescent systems, star-forming objects are heavily
affected by internal dust extinction
\citep{wyder,johnson07,driver07,luca,tojeiro09}. In such galaxies, UV
radiation heats dust grains which then re-radiate at far-infrared
(FIR) wavelengths. However, thermally emitting dust detected by, 
for example, {\it IRAS} (at $25-100\mu$m) can often only constitute
a small fraction of a galaxy's total dust mass.  Results from the
SCUBA Local Universe Galaxy Survey \citep{dunne01,vlahakis05}, 
as well as those of the sample of the Virgo cluster studied in \citet{Popescu02} 
and \citet{Tuffs02}, using the ISOPHOT instrument on board the {\it Infrared Space Observatory} ({\it ISO}),
show
that there is a population of galaxies containing much larger
proportions of cold dust that radiates at $>$ 100 $\rm{~\mu m}$ in
submillimeter bands.  As such, submillimeter observations provide
invaluable information for estimating the total dust content of these
galaxies which in turn helps us to understand relationships with other
galaxy properties. For instance, although we know there is a clear
division between the optical properties of galaxies on the red
sequence and those in the blue cloud, we do not know if this is
reflected in their submillimeter properties.
   
In the present study, we use submillimeter data acquired by the
{\it Herschel} Space Observatory as a part of the {\it Herschel} Astrophysical
Terahertz Large Area Survey \citep[H-ATLAS;][]{eales}.  The H-ATLAS survey 
is an open-time programme to survey $\sim$600 deg$^2$ of the
extragalactic sky over the wavelength range 110--500 $\rm{~\mu m}$
using the Photodetector Array Camera and Spectrometer
\citep[PACS;][]{pogli} and Spectral and Photometric Imaging REceiver
\citep[SPIRE;][]{griffin}. H-ATLAS also covers the largest area in a blind extragalactic survey with  {\it Herschel}.
One of the aim if to provide the FIR equivalent for the SDSS and 2dFGRS \citep[Two-degree Field Galaxy Redshift Survey;][]{colless01} surveys  is to characterize the properties of nearby galaxies. Also, for all regions covered by H-ATLAS, spectroscopic completeness will be guaranteed only up to the redshift of the GAMA survey \footnote{A detailed description of the GAMA survey is given in Sec.\ref{gamasurvey}}\citep[e.g. $z \lesssim 0.2$;][]{driver10}. So detailed statistical analysis will mainly be possible in the local universe. 
Therefore it is crucial to understand what kind of nearby galaxies are detected by H-ATLAS to provide a useful characterization of the FIR properties of local galaxies. In addition, we are also going to characterize the relationship between the nature and environment of
submillimeter sources detected in H-ATLAS. So by concentrating on sources at low redshifts, we are able to have a better estimation of the 
galaxy projected density since such measurements become difficult at higher redshifts due to  
a decrease in the observed limiting  luminosity of galaxies with spectroscopic redshifts. Also, nearby sources
can be spatially resolved far more easily (e.g. less blending), so that we can see where the submillimeter emission is coming from, unlike for sources at high redshifts. 

The submillimeter data analysed in
this study were acquired during {\it Herschel}'s Science Demonstration Phase
(SDP) and cover an area of $\sim 14.5$ deg$^2$.
Three main advantages of {\it Herschel} over previous infrared observatories
are its broad wavelength range, high sensitivity and high angular
resolution. The wavelength range in particular encompasses the peak of
the far-IR/submillimeter part of the spectral energy distribution for
low redshift galaxies thus allowing accurate measurements of total dust
mass.  In contrast, other submillimeter surveys such as those conducted by the {\it Balloon-borne Large-Aperture Submillimeter Telescope} ({\it BLAST})
with observations at 250$\rm{ \mu m}$, 350$\rm {\mu m}$, and 500$\rm{\mu m}$, have lower angular
resolution and do not cover more than $\sim$20 deg$^2$ of the extragalactic sky  \citep{pascale08,devlin}.

The layout of this paper is as follows. In \S2, we provide a summary
of our submillimeter sample selection, optical photometric and
spectroscopic observations from the SDSS/GAMA surveys, and UV
observations from GAMA/{\it GALEX} survey \citep{Seibert10}. A summary of the parameters estimated from
our data is given in \S3. Our final results and concluding remarks are
given in \S4.


\section{Data}
We will use the SDSS DR7 \citep{Abazajian09} together with the Galaxy And Mass Assembly redshift survey \cite[GAMA
survey;][]{driver10} to compile an optically selected sample of low-redshift galaxies, 
and from there, look at the optical properties of those sources detected by {\it Herschel}. Here, we
summarize the data in each survey and describe our sample selection.

\subsection{Optical/UV data}
\label{gamasurvey}
 
Optical sources used in this paper were initially 
taken from the SDSS DR7 \citep{Abazajian09}.  Our aim is to have an
optically selected sample of galaxies with the criterion $0.02 \leq z
\leq 0.2$.  To do this, all SDSS sources from the \texttt{GALAXY} table were selected and
their positions were cross-matched within 0.5 arcsec with objects from
the GAMA survey in order to extract their spectroscopic
redshifts as well as colours.  The GAMA survey is based on data from SDSS DR6 and
UKIDSS-LAS DR4 and consists of three equatorial regions named 
GAMA-09h, GAMA-12h, and GAMA-15h.  The field observed  by {\it Herschel},
coincides with the GAMA-09h equatorial region ($\rm{Ra=129.0,141.0~deg ; Dec=-1.0, +3.0~deg}$) 
where its spectroscopic survey is complete up to Petrosian magnitude, $r\la19.4$ \citep{aaron}. 
Moreover, the GAMA data contains $r$-band defined aperture matched photometry in FUV/NUV (far-ultraviolet/near-ultraviolet) from {\it GALEX} and in {\it ugriz} optical bands from
SDSS \citep{hill11}. 
We only select objects with the
GAMA spectroscopic redshift quality parameter $Q\geq3$, which approves
their inclusion in scientific analysis \citep{driver10}. In our optical sample, $\sim$1.4~per cent of sources with $0.02 \leq z
\leq 0.2$ have $Q<3$.

We find a total of $\sim 3370$ galaxies in the redshift range
$0.02 \leq z \leq 0.2$ within $\approx$12.5 deg$^2$ of our
$\approx$14.5 deg$^2$ SDP field. The coverage is not complete,
because the GAMA data goes down to 
$\rm{Dec=-1.0~deg}$.  In addition to optical fluxes, NUV magnitudes were available
for $\sim 2680$ galaxies \citep{Seibert10,hill11}. All magnitudes are corrected for 
Galactic extinction.

\subsection{Submillimeter  data}

The H-ATLAS SDP data were acquired in November 2009 using {\it Herschel}'s
parallel/fast scan mode, scanning at 60 arcsec s$^{-1}$. In this mode,
mapping by PACS at 100 and 160 $\rm {\mu m}$ and SPIRE at 250, 350,
and 500$\rm {\mu m}$ were carried out simultaneously. The
SDP data covers an area of $\sim$ 4.0~deg $\times$ 3.6~deg
centered on $(\alpha,\delta) \approx (09^{\rm h}05^{\rm m}, +0^{\rm
\circ}30^{\rm '})$.
 
A complete description of the data reduction and noise analysis of
PACS and SPIRE data are given in \citet{ibar} and \citet{pascale}, respectively.  Observed time-line data from both instruments were
processed by using the {\it Herschel} Interactive Processing Environment
(HIPE) using custom reduction scripts.  Thermal drift in bolometer
arrays was corrected by applying high-pass filtering. The naive map-making
method of HIPE was used to project the two cross-scan timeline
observations.
 
The estimated point spread functions (PSFs) of the final maps have
full width at half maximum (FWHM) values of 8.7, 13.1, 18.1, 25.2, and
36.6 arcsec at 100, 160, 250, 350 and 500 $\rm {\mu m}$, respectively. The estimated 1$\sigma$ noise levels at 100, 160, 250,
350, and 500$\rm {\mu m}$ are 25--30, 33--48, 6.6 , 7.6, and 9.0
mJy/beam, respectively \citep{ibar,rigby}. 
 
Sources were extracted using the 250 $\rm {\mu m}$ map as described in
\citet{rigby}.  The typical positional error for a $\geq$5~$\sigma$ source is 2.5 arcsec or less.
For each 250 $\rm {\mu m}$ source, corresponding 350
and 500 $\rm {\mu m}$ flux densities were measured by using the
noise-weighted/beam-convolved 350 and 500 $\rm {\mu m}$ maps at the
source position determined from the 250 $\rm {\mu m}$ map.  Fluxes at
100 and 160 $\rm {\mu m}$ were estimated by matching each 250 $\rm {\mu m}$
source to the nearest PACS sources within a radius of 10 arcsec.

\subsubsection{The submillimeter source catalogue}

The H-ATLAS standard source catalogue of \citet{rigby} is supplemented
with cross-identification information from GAMA and SDSS DR7 surveys
as described in \citet{smith}. A likelihood-ratio analysis \citep{sutherland} is performed by \citet{smith} to match 250$\rm {\mu m}$ sources to SDSS DR7 sources brighter than $r$=22.4~mag.
The probability that an optical source is in fact associated with the submillimeter source has been used to define the \texttt{Reliability} parameter.
According to \citet{smith}, objects with \texttt{Reliability}$\geq 0.8$ are considered to be true matches to submillimeter sources.

While we use the SDSS galaxy IDs to obtain
the H-ATLAS PACS and SPIRE submillimeter fluxes, we apply the same
reliability cut as suggested by \citet{smith}, i.e. \texttt{Reliability}$\geq 0.8$ which guarantees
that we are $\sim$93 per cent ($\sim$ 83 per cent) complete in the redshift range
$0.02 \leq z \leq 0.1$ ($0.1 < z \leq 0.2$).  As such, throughout this
paper, we refer to a 'detected source' as an SDSS galaxy with a $\geq
5.0 \sigma$ submillimeter counterpart that has
\texttt{Reliability}$\geq 0.8$.  Accordingly, we detect 496 galaxies
at 250$\rm {\mu m}$ of which 482 ($\sim$ 97.2 per cent) have associated NUV
fluxes. Thus in 14 galaxies ($\lesssim$ 2.8 per cent) with submillimeter detections we found no UV counterpart, but this will not affect the main results of the paper. 


\section{Analysis}
\subsection{Summary of source detection}
\label{statistics}

The first step of our analysis is to determine the rate of source
detection in H-ATLAS as a function of galaxy $r$-band magnitude.
The $r$-band magnitude is useful since it can be used as a starting
point in sample selection for other extragalactic studies of low
redshift galaxies using the {\it Herschel} submillimeter data. To do so, for all
galaxies,  we plot $(u-r)$ vs.
Petrosian magnitude $r_{_{\rm Petro}}$ corrected for Galactic
extinction \footnote{The SDSS Petrosian magnitude  is a modified form of the \citet{Petrosian} system in which galaxy fluxes are measured within a circular aperture ${\rm \mathcal{R}_P}$(Petrosian radius) such that the ratio of the local surface brightness in an annulus at ${\rm \mathcal{R}_P}$ to the mean surface brightness within ${\rm \mathcal{R}_P}$, is equal to some constant value. Alternatively we can use the SDSS Model magnitude $r_{_{\rm Model}}$ though it does not change the results since from a sample of 10000 galaxies within the magnitude/redshift range of our sample we find 
$r_{_{\rm Petro}} \approx r_{_{\rm Model}} $+0.07.}.

\begin{figure*}
\includegraphics[width=18cm]{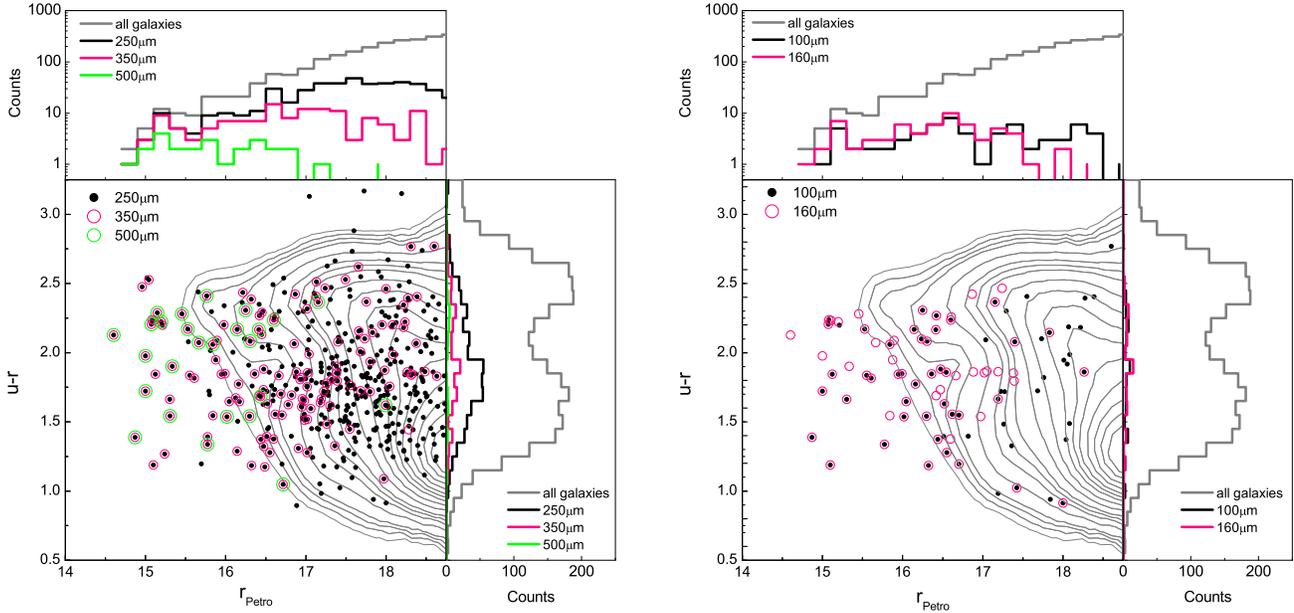}
\caption{Colour-magnitude diagram of optically selected galaxies (grey
  contour lines) and their submillimeter counterparts with
  \texttt{Reliability}$\geq$0.8 detected at $\gtrsim5\sigma$ by SPIRE
  (left panels) and PACS (right panel). Horizontal and vertical
  histograms show the distribution of galaxy optical colour
  ($u-r$) and $r$-band magnitude ($r_{_{\rm Petro}}$), respectively. Colours represent: 'grey' all galaxies; 'black' sources with SPIRE 250$\rm{ \mu m}$ (left panel) and PACS 100$\rm{ \mu
    m}$ (right panel) detections; 'pink' sources with SPIRE 350$\rm{ \mu m}$ (left panel)
  and PACS 160$\rm{ \mu m}$ (right panel) detections; 'green' sources with
  SPIRE 500$\rm{ \mu m}$ detections. }
\label{NUV-r}
\end{figure*}

As Fig.~\ref{NUV-r} shows, the overall distribution of galaxies (gray
contour lines) in $u-r$ colour is bimodal, showing blue cloud and red
sequence galaxies.  The overlaid data points represent galaxies with
submillimeter detections made by SPIRE (left panel) and PACS (right
panel).  Horizontal and vertical histograms in each panel show the
distribution of data points corresponding to individual SPIRE/PACS bands
along the magnitude and colour axes, respectively.

Fig.~\ref{NUV-r} shows that the majority of galaxies with
submillimeter detections are located in the blue cloud, with histograms
of their colour peak around $u-r$$\approx$1.4 without substantial
red tails. Thus, within the redshift slice of the current sample of
galaxies, the survey on the whole is mapping the blue
sequence. Furthermore, a comparison between histograms at 250 $\rm
{\mu m}$ and those in other SPIRE/PACS bands show that the SPIRE 250
$\rm {\mu m}$ band is by far the most sensitive, since the estimated
rate of detections are higher at 250$\rm{ \mu m}$ compared to other
submillimeter bands. This is due to the higher sensitivity of SPIRE
over PACS as well as the higher resolution of the 250$\rm{ \mu m}$
map. With this in mind, in the analysis that follows, we only consider
sources with 250 $\rm {\mu m}$ detections.

\begin{figure*}
\includegraphics[width=18.5cm]{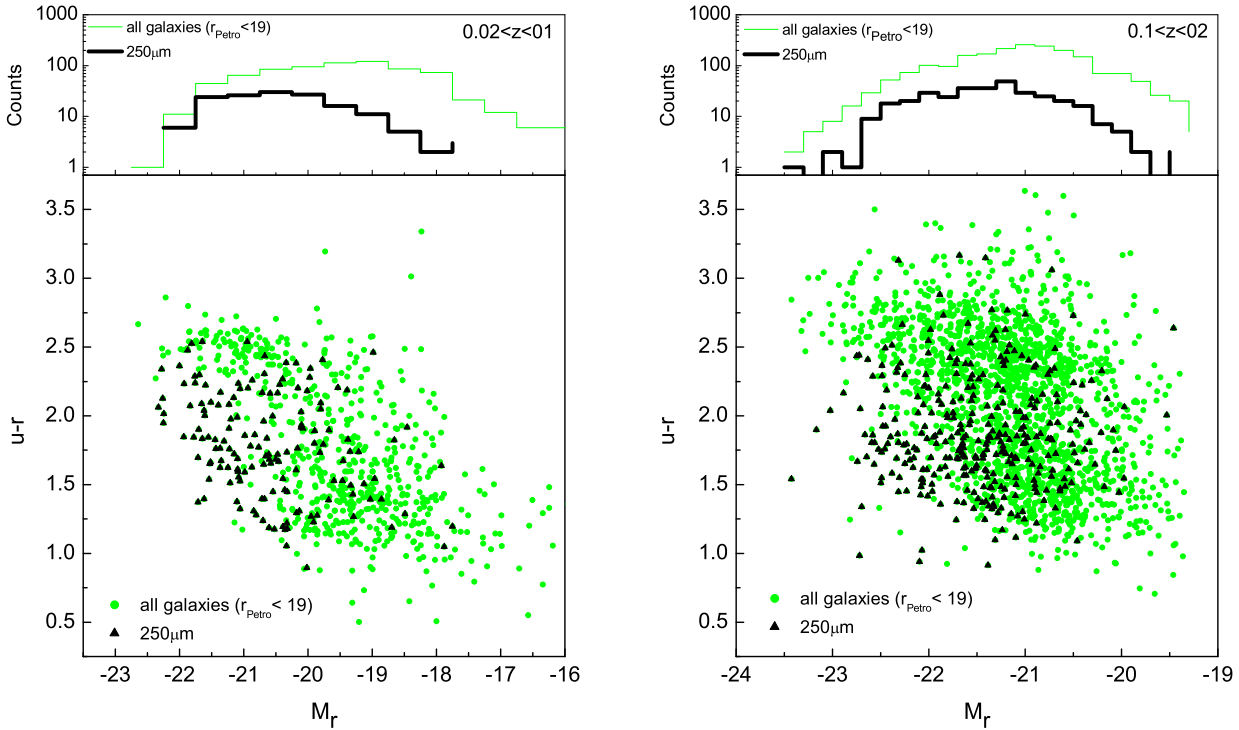}
\caption{Same as in Fig.~\ref{NUV-r} but as a function of absolute
  $r$-band magnitude M$_{\rm r}$ and for 250$\rm{ \mu m}$ detections. 'Left' and 'right' panels correspond to redshift bins $0.02
  \leq z \leq 0.1$ and $0.1 \leq z \leq 0.2$, while horizontal
  histograms show the distribution of galaxy $u-r$ colour
   and absolute $r$-band magnitude, respectively. Colour of data-points/histograms represent: 'green' all galaxies; 'black' sources with SPIRE 250$\rm{ \mu m}$  detections.}
\label{NUV-r2}
\end{figure*}

The quantities plotted in Fig.~\ref{NUV-r} are also shown in
Fig.~\ref{NUV-r2}, but as a function of absolute $r$-band magnitude
$M_r$ and for sources with 250 $\rm {\mu m}$ detections. 'Left' and
'right' panels correspond to redshift bins $0.02 \leq z \leq 0.1$ and
$0.1 \leq z \leq 0.2$, respectively.  The absolute magnitude $M_r$ is
given by 
\begin{equation}
M_r = r_{_{\rm Petro}} - k_r - 5\log \left( \frac{D_{\rm {L}}}{10{\rm pc}} \right), 
\end{equation}
where $D_{\rm{L}}$ is the luminosity distance, and $k_r$ is the k-correction
using the method of \citet{Chilingarian}. The adopted cosmology to
estimate $D_{\rm{L}}$ is a flat Universe with the matter density parameter
$\Omega_{\rm{M}}$=0.3 and cosmological constant $\Omega_{\Lambda}$=0.7, with
$H_0$=70~km s$^{-1}$ Mpc$^{-1}$.  The 'horizontal green histograms' in Fig.~\ref{NUV-r2} show that the rate of detections of galaxies in the $r$-band  become incomplete for sources with $M_r \gtrsim -21.0$~mag and $M_r
\gtrsim -21.5$~mag in the redshift bins $0.02 \leq z \leq 0.1$ and $0.1
\leq z \leq 0.2$, respectively with almost double the number of
detections in the higher redshift slice. Thus in the low-redshift Universe and within the current depth of H-ATLAS, most of the {\it Herschel} detections in the $r$-band consist of more luminous galaxies (and
therefore those with larger stellar masses) which have lower values of $u-r$.

In order to analyze our sample in more detail, we divide galaxies into
two populations by applying a cut on their $u-r$ colour indices.  To
find an optimum colour cut, a double-Gaussian function is fitted to
the histogram for colour of all galaxies with
$r_{\rm{_{Petro}}}\lesssim19.0$~mag.  The combined function has a minimum around
$u-r\approx2.2$~mag. As such, from here on, unless otherwise stated, we refer to sources with
$u-r\leq2.2$~mag and $u-r>2.2$~mag as {\it blue} and {\it red}
objects, respectively.

\begin{figure*}
\includegraphics[width=7.5cm]{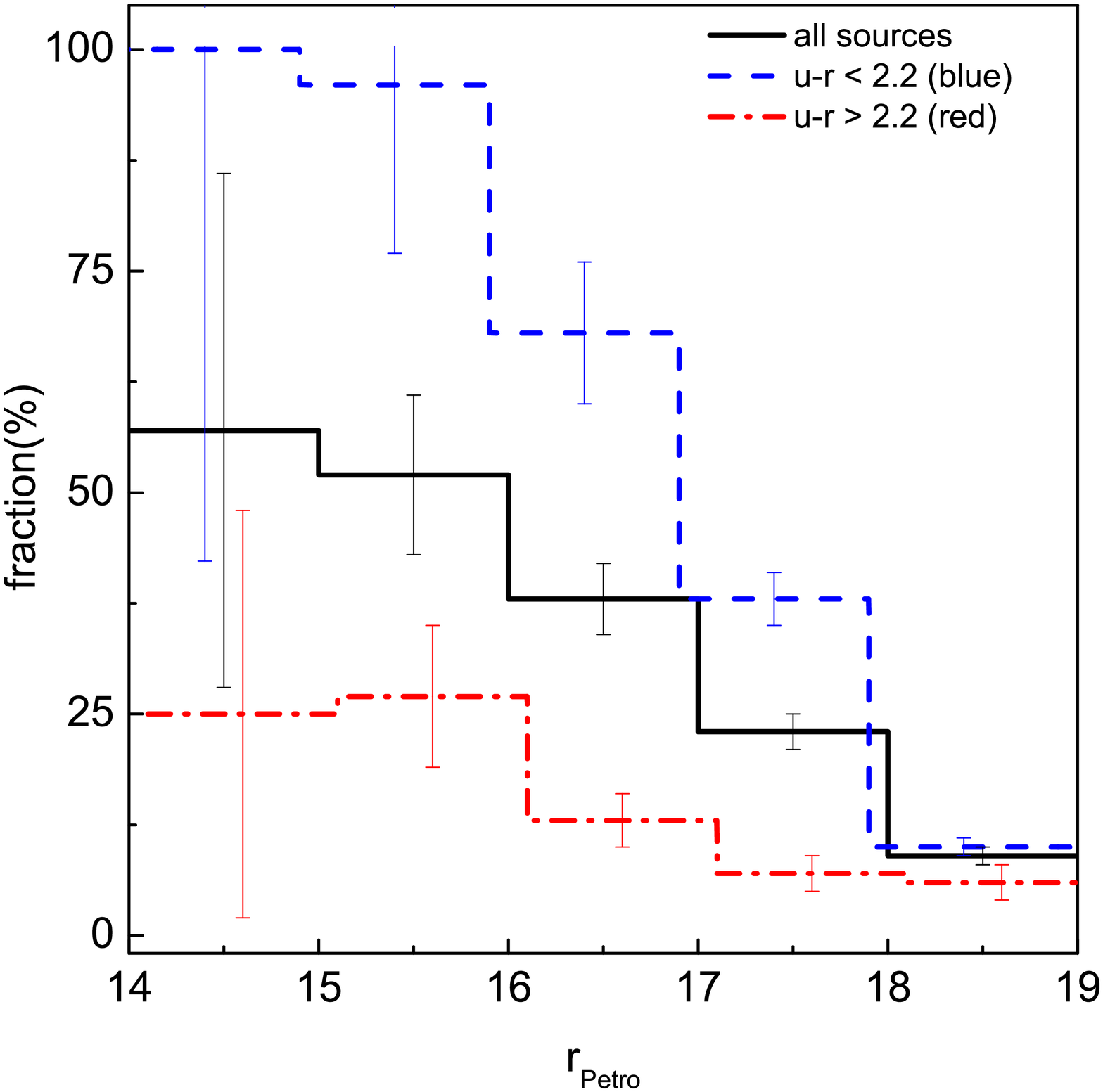}
\caption{Histograms of the fraction of detected galaxies as a function
  of $r$-band magnitude. The black histogram includes all galaxies, while
  the two other histograms represent the detection rate of blue (blue
  dashed histogram; $u-r \leq$2.2) and red (red dash-dotted histogram;
  $u-r >$2.2) sources, respectively.  }
\label{detection}
\end{figure*}

In Fig.~\ref{detection} we examine the differential fraction of detections of galaxies at 250$\rm{ \mu m}$ as a function of $r$-band magnitude. 
We do this using $\sqrt{n}$ counting errors ($n$ is the number of detected sources in each magnitude bin) for all optical sources (black histogram) as well as blue (blue dashed histogram) and red (red dash-dotted histogram) galaxies.
The results, as presented in Fig.~\ref{detection}, show that
(i) In general $\gtrsim 50$ per cent of SDSS galaxies with $r_{_{\rm Petro}} \leq
17.0$~mag have 250$\rm{ \mu m}$ counterparts in H-ATLAS.
(ii) The submillimeter detection rate falls off rapidly as one moves toward
fainter galaxies. This is mainly due to a decrease in the observed
signal-to-noise ratio toward fainter galaxies.  (iii) In each bin of
$r$-band magnitude and for $r_{_{\rm Petro}} \lesssim 18.0$~mag, the
detection fraction is $\gtrsim 4$ times higher for 'blue' galaxies
when compared to the `red' population, indicating that the vast majority
of detected sources are blue objects.  Although such blue sources are
in general star-forming galaxies, it does not mean that 250$\rm{ \mu
m}$ red detected objects are passive galaxies. We will discuss the
properties of such sources in Section~\ref{ccplot}.

Finally in Table.~\ref{stackTable} we present the total percentage of
sources detected by H-ATLAS in samples of SDSS galaxies with differing
limiting magnitudes. Samples of SDSS galaxies in the second column
include galaxies with all colour indices, while those in the third and
fourth columns contain blue and red objects, respectively.


\begin{table*}
  \caption{The total fraction of sources detected by H-ATLAS among
    SDSS galaxies to differing limiting magnitudes, as shown in the
    first column. Samples of SDSS galaxies in the second column
    include galaxies with all colour indices, while those of third and
    forth columns contain blue and red objects, respectively.}
  \label{stackTable}
  \smallskip
    \begin{tabular}{|l|c|c|c|}
      \hline\hline
          sample limiting magnitude &      &   Total fraction detected (percent)   &     \\  
         & all colours & blue ($u-r\leq2.2$~mag)   & red($u-r>2.2$~mag) \ \\  \hline\hline
     $r_{_{\rm Petro}} \leq 16.0$~mag & $53\pm8$ & $97\pm18$ & $27\pm8$  \\    
     $r_{_{\rm Petro}} \leq 17.0$~mag & $42\pm4$ & $75\pm7$  & $17\pm3$  \\
     $r_{_{\rm Petro}} \leq 18.0$~mag & $29\pm2$ & $47\pm3$  & $11\pm2$   \\
     $r_{_{\rm Petro}} \leq 19.0$~mag & $17\pm1$ & $23\pm2$  & $9\pm1$   \\  \hline
         \end{tabular}\par
   \vspace{.5\skip\footins}
\end{table*}


 \subsection{Environment of detected sources}
 \label{sig5}
 
 \begin{figure*}
\includegraphics[width=18.5cm]{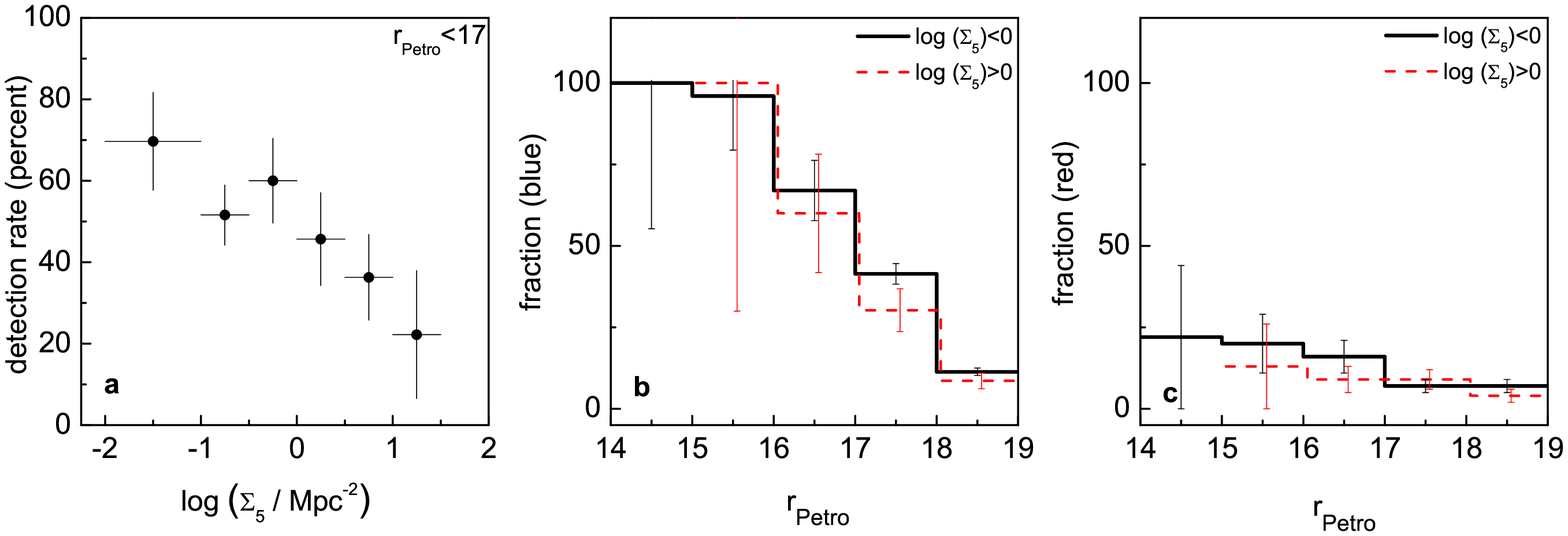}
\caption{{\bf (a)} The fraction of SDSS galaxies detected in H-ATLAS
  as a function of galaxy local density $\Sigma_5$ estimated using
  Eq.~\ref{density22}.  {\bf (b)} Histograms of the fraction of
  detected blue ($u-r \leq$2.2~mag) galaxies as a function of
  $r$-band magnitude. 'The red dashed histogram' includes galaxies in high
  density regions ($\log (\Sigma_5/{\rm Mpc^2}) \geq 0$) and 'the black
  histogram' includes those in low density regions ($\log
  (\Sigma_5/{\rm Mpc^2}) < 0$). {\bf (c)} Same as in 'b' but for red
  ($u-r >$2.2~mag) galaxies.  }
\label{detection2}
\end{figure*}

Our results in \S\ref{statistics} show that submillimeter detected
sources are preferentially blue sequence galaxies. Thus, since blue
sequence galaxies are much more frequent in low density environments,
we expect to find a lower detection rate in group/cluster environments
than in the field.

To explore the environmental density of submillimeter detected
galaxies, we consider the projected surface density, based on counting
the number of nearest neighbours, i.e. the density within the distance
to the Nth nearest neighbour.  Although this is a 2D estimate, the
redshift information of each galaxy is used to remove the background
and foreground sources. The environmental density $\Sigma_N$ around
each galaxy detected at 250$\rm{ \mu m}$ is determined from the
following relation:
\begin{equation}
\Sigma_{N} ({\rm Mpc^{-2}})=\frac{N}{\pi d^2_{N}},
\label{density22}
\end{equation}
where $d_N$ is the projected comoving distance to the $N^{\rm{th}}$ nearest
neighbour that is within the allowed redshift range $\pm \Delta z
c$=1000~km~s$^{-1}$ \citep{balogha,Baldry06}. We set $N=5$ in our
analysis. The neighbouring galaxies are those with spectroscopic
redshifts and $M_r \leq M_{_{\rm r,limit}}-Q(z-z_0)$, where $Q$=1.6 is
the evolution of the galaxy $r$-band luminosity relative to $z_0 =
0.1$, determined by \citet{blanton03}.  Horizontal histograms (grey
thick lines) in the 'left' and 'right' panels of Fig.~\ref{NUV-r2} show
that galaxy counts become incomplete for the $M_r \gtrsim-19.0$~mag range if
$0.02 \leq z \leq 0.1$ and $M_r\gtrsim$-20.5~mag if $0.1 \leq z \leq
0.2$. Thus the parameters ($M_{_{\rm r,limit}}$,$z_0$) were set to
$(-20.5,0.1)$.

The rate of detection of the H-ATLAS sources as a function of the density
parameter $\Sigma_5$ is plotted in panel 'a' of
Fig.~\ref{detection2}. Error bars are Poisson errors on the
mean. Detection rates have been estimated for galaxies with $r_{_{\rm
Petro}} \leq 17.0$~mag, since galaxies with larger $r$-band
magnitudes are not systematically detected due to their low
signal-to-noise ratios. In other words, the majority of fainter galaxies are
not detected in H-ATLAS, irrespective of their environments.

It is evident that the fraction of detected sources in low-density
environments is larger by a factor of two in comparison to the rate of
detection in high-density environments. This is expected, since the
majority of detected objects in H-ATLAS are blue galaxies that
preferentially populate the low-density regions. Results remain more
or less the same if we use $\Sigma_4$ or $\Sigma_6$ instead of
$\Sigma_5$, showing that the dependence of the detection rate on 
environmental density is not sensitive to the choice of $N$ in
Eq.~\ref{density22}.  To obtain a better understanding of the type of
detected objects that populate either low or high density environments, we
plot the rate of detection of galaxies as a function of $r_{_{\rm
Petro}}$ for sources with different colours and densities in
Fig.~\ref{detection2} (see panels 'b' and 'c').

The estimated projected density ranges from $-2.0$ to $2.0$ on a
logarithmic scale. We therefore define the threshold between low and
high density environments at $\log \Sigma_5 \simeq 0$. As such, in
panels 'b' and 'c' of Fig.~\ref{detection2}, black histograms
correspond to sources in low density environments with $\log
(\Sigma_5/{\rm Mpc^2}) < 0 $ while red dashed histograms refer to
the fraction of detected galaxies in high density regions, i.e. $\log
(\Sigma_5/{\rm Mpc^2}) \geq 0 $. In addition, sources in panel 'b'
have index $u-r \leq$~2.2~mag (e.g. blue objects) and the colour of
those in panel 'c' is $u-r >$~2.2~mag (e.g. red objects). Although
the number of blue galaxies is smaller in high-density regions, panel
'b' shows that the rate of detection remains the same in either
environment. The situation is similar for the red galaxies, although
the overall rate of detection of red objects is much smaller than for blue
ones.


\subsection{Nature of RED 250$\rm{\mu}m$ sources}
\label{ccplot}

\begin{figure*}
\includegraphics[width=18cm]{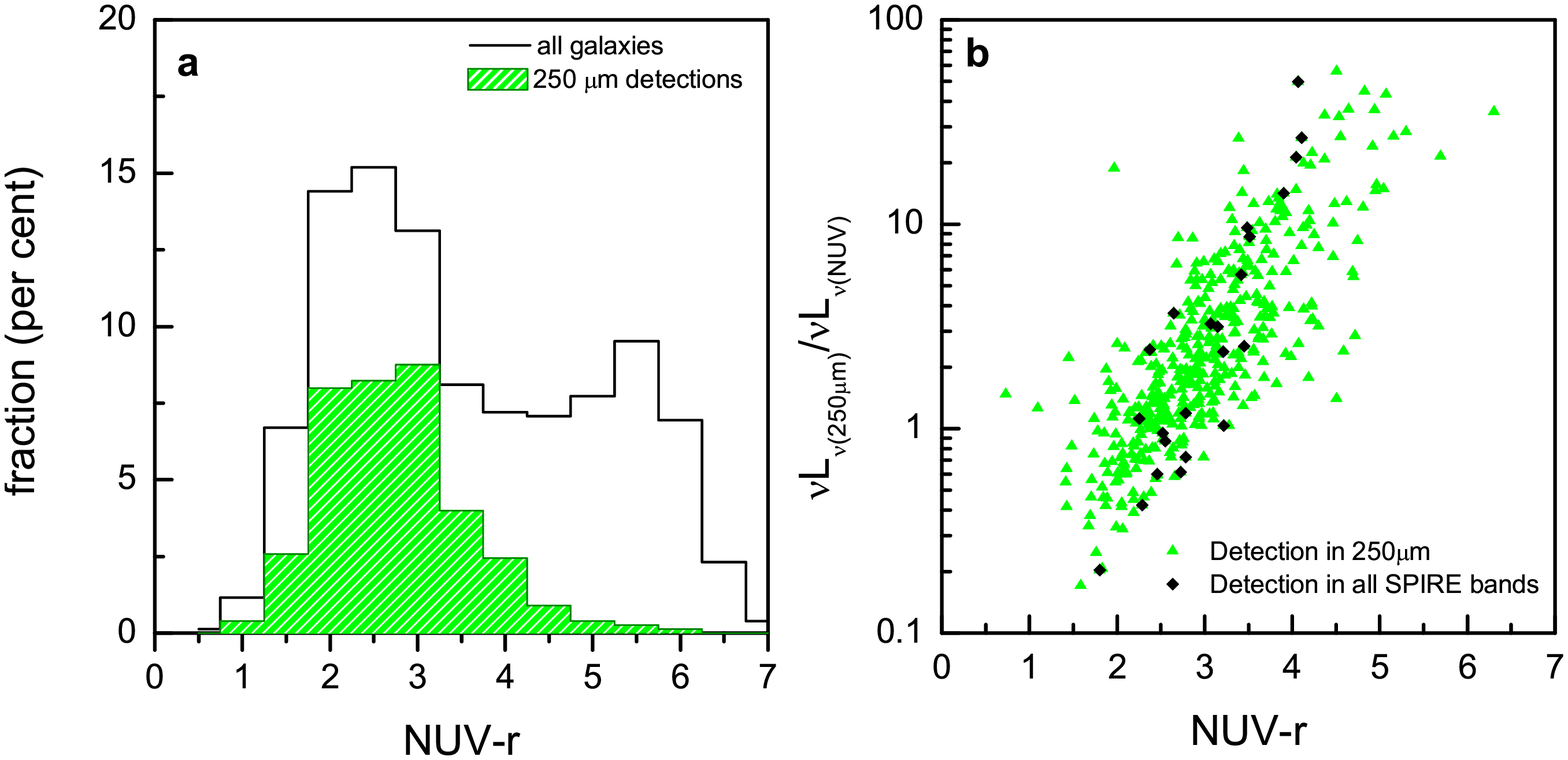}
\caption{{\bf (a)} NUV$-r$ colour distributions for all the SDSS galaxies (black histogram) with available UV fluxes as well as those detected in 250$\rm{\mu}m$ (green-shaded histogram).{\bf (b)}The NUV$-r$ vs. $\nu L_{\rm{250}}$/$\nu L_{\rm{NUV}}$ for galaxies with 250$\rm{\mu}m$ detections (green triangles). Overlaid (black
  diamonds) are galaxies detected in all SPIRE bands.}
\label{NUVcolour}
\end{figure*}

We have already seen that the majority ($\sim$~80 per cent) of galaxies detected in H-ATLAS at low redshift are blue galaxies while the rest $\sim$~20 per cent are red. However, such a fraction on red detected sources is an upper limit since many of these galaxies might have traces of recent star formation activities which is not reflected in their optical $u-r$ colour. In fact, UV-optical colours are more robust in classifying galaxies into blue/red categories in comparison to optical colours since they are more sensitive to recent star-formation activity \citep[e.g.][]{Schawinski07,Cortese09}. The reason that we did not use UV-optical colour in Sec.3.1 is that for $\sim$~20 per cent of SDSS galaxies in the main sample, UV fluxes were not available from GAMA (see Sec.2.1). These galaxies (e.g. those without UV counterpart) are basically faint while their optical colour distribution peaks around $u-r\approx$~2.5. So in order to avoid any biases in analyzing the fraction of source detection in previous sections, we were restricted to use  the SDSS optical colours. Fortunately, NUV fluxes are available for $\gtrsim$~98 per cent of detected sources in 250$\rm{\mu}m$ which enable us to investigate the spectral properties of galaxies with 250$\rm{\mu}m$ detections from their UV-optical colours.

Thus in Fig.~\ref{NUVcolour}a, the NUV$-r$ colour distributions for all the SDSS galaxies  with available UV fluxes (black histogram) as well as those detected in 250$\rm{\mu}m$ (green-shaded histogram) are shown. The black histogram in Fig.~\ref{NUVcolour}a shows a bimodality with a minimum around NUV$-r\approx4.5$~mag found from fitting a double-Gaussian function to it. In addition, $\sim$~95 per cent of detected sources have NUV$-r\leq$~4.5~mag (i.e., blue in UV-optical colour) and just a small population of detected sources have  NUV$-r\geq$~4.5~mag (i.e., red in UV-optical colour). 
 
It is therefore interesting to know whether such red objects (i.e., NUV$-r\geq$~4.5~mag) are
obscured star-forming systems or whether they are a
population of quiescent/passive galaxies.  As such, for all detected galaxies, a UV-optical vs. infrared
colour-colour diagram is plotted in Fig.~\ref{NUVcolour}b, where green
triangles represent sources detected at 250$\rm{\mu}m$ and
highlighted data points (black diamonds) represent those H-ATLAS sources detected in all three SPIRE bands.
Not surprisingly, the plot shows a correlation between NUV$-r$ and
$\nu L_{\rm{250}}$/$\nu L_{\rm{NUV}}$, with red galaxies having a higher
$\nu L_{\rm{250}}$/$\nu L_{\rm{NUV}}$ ratio than blue systems. A similar
relation was found by \citet{johnson07} between
the total far-infrared-to-near-ultraviolet luminosity ratio 
($L_{\rm{TIR}}$/$L_{\rm{NUV}}$) and NUV$-r$ for a sample of local
galaxies selected from SDSS.

By combining this relation with the library of infrared spectral energy 
distributions of \citet{Chary01}, we can investigate the real 
nature of the H-ATLAS sources with NUV$-r>$4.5. 
According to \citet{Chary01}, the $\nu L_{\rm{250}}$/$L_{\rm{TIR}}$ luminosity 
ratio varies in the range $\sim$4-65. Even assuming the lowest possible value (i.e., $\sim$4), 
this implies a $L_{\rm{TIR}}$/$L_{\rm{NUV}}$ ratio $\sim$6-250 for our red galaxies and 
we can use this value to estimate their NUV dust attenuation $A(NUV)$.
If we adopt the typical relation between $L_{\rm{TIR}}$/$L_{\rm{NUV}}$ and $A(NUV)$ 
for star-forming objects \citep[e.g.][]{buat05}, we find $A(NUV)\sim$1-4 mag. 
Interestingly, even using the relations between 
$L_{\rm{TIR}}$/$L_{\rm{NUV}}$ and $A(NUV)$ presented by \citet{luca} 
which take into account the heating of dust coming from 
evolved stellar populations, we still obtain $A(NUV)\geq$0.2-1.8\footnote{These values 
have been obtained by using the most extreme case discussed 
by \citet{luca}, i.e. $\tau<2.8$, and they are thus 
lower limits to the real values for our sample.}.

Thus, this simple exercise shows that not surprisingly, 
the vast majority ($\geq$90 per cent) of the red objects detected by H-ATLAS in our 
sample have 'corrected' NUV$-r$  colour lower than 4.5 mag, i.e., they are 
red not because they are old/passive but mainly because they are 
obscured by dust. Of course, the presence (even if low in fraction)
of a population of 'truly passive' systems in the H-ATLAS survey is a very
intriguing possibility and might have important implications for our understanding
of dust properties in galaxies. A detailed search for FIR emitting 'passive' systems and a discussion of their
properties will be presented in a future work \citep{kate}.


\subsection{Variation of temperature with density}
\label{temp-density}

Our findings in Section~\ref{sig5} show that, irrespective of their
environments, the rate of detection of submillimeter sources remains
the same in either high or low density regions.  Therefore it might be
interesting to test whether the submillimeter properties (e.g., colours
and dust temperature) of detected sources in H-ATLAS correlate with
the environment.

Out of 71 galaxies with $\geq3\sigma$ detections in all SPIRE bands,
38 have associated ($\geq3\sigma$) PACS detections.  To estimate the
dust temperature of these 38 galaxies, we fit the submillimeter
photometry with a single component modified black-body SED over the
wavelength range 100--500$\rm{ \mu m}$.  The equation for a single
component black body is given by
\begin{equation}
F_{\nu} = \frac{\kappa_{\nu}}{D^2_{\rm{L}}} M B_{\nu}(T).
\end{equation}
Here $M$ is the dust mass, $T$ is the dust temperature, $B_{\nu}(T)$ is
the Planck function, $D_{\rm{L}}$ is the luminosity distance to the galaxy,
and $\kappa_{\nu}$ is the dust emissivity. The dust emissivity is assumed to 
be a power law in this spectral range, where $\kappa_{\nu}\propto\nu^{\beta}$. 
As we are not constraining dust masses, an absolute value of $\kappa$ is not needed. 
We assume $\beta = 1.5$, though we note that the value is notoriously
uncertain. 

The uncertainties in the flux densities adopted from the standard H-ATLAS catalogue
\citep[e.g.][]{rigby} include the contribution from instrumental and
confusion noise only. Thus we add the calibration errors of 10--20 per cent in the
 100--160$\rm{ \mu m}$ PACS bands and 15 per cent in all SPIRE bands in
quadrature to those given in the H-ATLAS catalogue.
The best-fit solution is found by minimizing the
chi-squared ($\chi^2$) function, yielding an average uncertainty in the
fitting of $\pm1.8$~K for temperature. The average reduced $\chi^2$
value for all the fits is 0.7, with a standard deviation of 0.6.
We find that the data in the 100--500$\rm{ \mu m}$ range are
accurately fitted by a single black-body component.  

\begin{figure*}
\includegraphics[width=17cm]{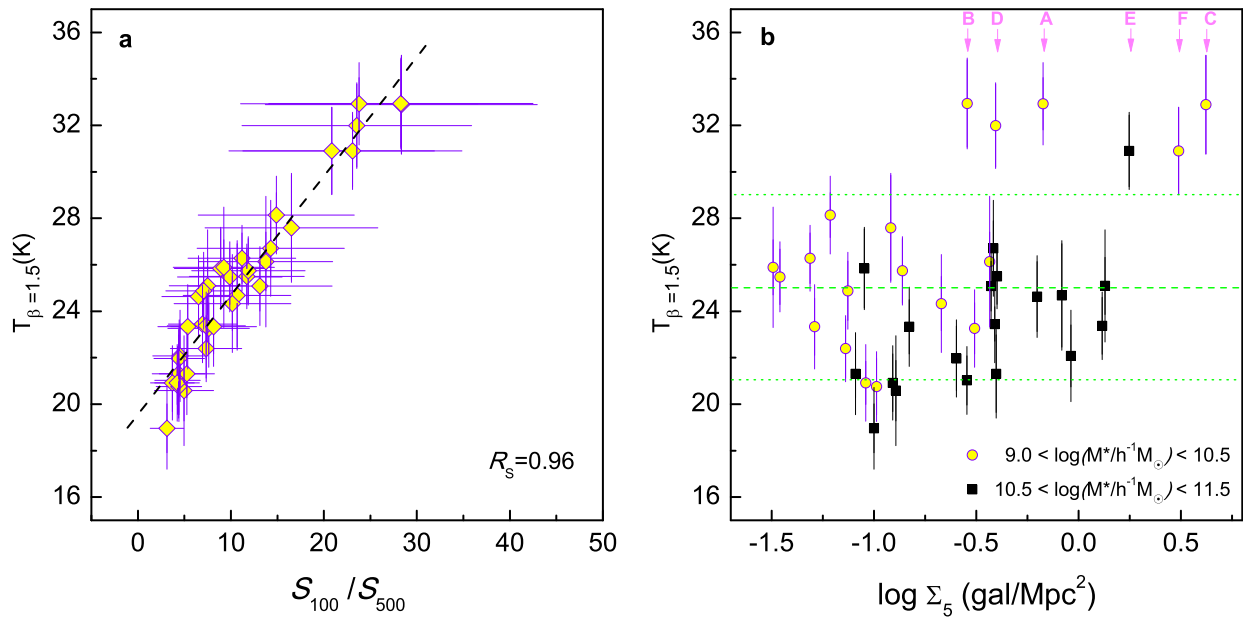}
\caption{{\bf (a)} Dust temperature vs. submillimeter colour
  $S_{\rm{100}}/S_{\rm{500}}$. The dashed line shows the linear
  regression fit as described by Eq.~\ref{correlation}.  {\bf (b)}
  Dust temperature as a function of the local density of
  galaxies, $\Sigma_5$, for sources with $\geq3\sigma$ detections in
  SPIRE/PACS bands.  The green dashed line intersects the y-axis at
  the mean temperature of the whole sample ($T\approx25$~K)
  while green dotted lines represent the standard deviation of the mean temperature.
  Circles and squares represent galaxies with low
  ($9.5\leq\log(M_*/{\it h}^{-2}M_{\odot})\leq 10.5$) and high
  ($10.5\leq\log(M_*/{\it h}^{-2}M_{\odot})\leq 11.5$) stellar
  mass, respectively. The SDSS images of objects marked as 'A' to 'F'
  (with 'A' having the maximum temperature) are shown in
  Fig.~\ref{SDSSimages2}.}
\label{TempDensity}
\end{figure*}

Fig.~\ref{TempDensity}a shows that the submillimeter colour
$S_{\rm{100}}/S_{\rm{500}}$ correlates with the dust temperature.  Such a
correlation also exists between the dust temperature and other
submillimeter colour indices, e.g., $S_{\rm{160}}/S_{\rm{350}}$,
$S_{\rm{250}}/S_{\rm{500}}$, etc.  However, not surprisingly, the relation
between the $S_{\rm{100}}/S_{\rm{500}}$ flux ratio and temperature is the
strongest one.  The estimated Spearman correlation coefficient in panel
'a' is $R_{\rm{S}}=0.96$ and the best-fit linear regression to the data
points (black dashed line) suggests the following relation for the
dust temperature:
\begin{equation}
T(K)=0.51(\pm0.02)\times S_{\rm{100}}/S_{\rm{500}}+19.5(\pm0.3).
\label{correlation}
\end{equation}

A plot of the estimated dust temperature as a function of the density
parameter $\Sigma_5$ for all galaxies in our sample is presented in
panel 'b' of Fig.~\ref{TempDensity}.  Galaxies are divided into two
subsamples based on their stellar mass content with circles and
squares corresponding to sources with low ($9.5\leq\log(M_*/{\it
h}^{-2}M_{\odot})\leq 10.5$) and high ($10.5\leq\log(M_*/{\it
h}^{-2}M_{\odot})\leq 11.5$) stellar masses, respectively.
Galaxy stellar masses, $M_*$, have been computed in units of
solar mass from the relationship given by \citet{yang} as follow:
\begin{equation}
    \begin{split}
     \log(M_*/{\it h}^{-2}M_{\odot})= -0.406+1.097(g-r)\\ 
          - 0.4(M_r-5\log {\it h}-4.64),
    \end{split}
  \end{equation}
using the $(g-r)$ colour and where $M_r$ is the 
$r$-band absolute magnitude corrected for extinction and redshift.
The green dashed line shows the mean temperature of the
whole sample ($T\approx25\pm4$~K)\footnote{The estimated mean temperature is $T_{mean}\approx22\pm4$~K in the case of $\beta = 2.0$. In fact our data show a strong correlation ($R_{\rm{S}}=0.99$) between the estimated temperatures based on either values of $\beta$ such that $T_{\beta = 2.0}=0.78(\pm0.01)\times T_{\beta = 1.5}+2.1(\pm0.2)$.}, while green dotted lines
indicate the standard deviation. This result fairly agrees with the rest-frame dust temperatures measured  for a sample of the H-ATLAS galaxies studied in \citet{Dye10} and \citet{Amblard10} as well as those of the sample of BLAST sources determined by \citet{Dye09}.
We note that the densities in panel 'b' of Fig.~\ref{TempDensity} span
values typical of field galaxies and poor groups, whereas we are not
able to trace very high density environments, such as compact groups
or clusters of galaxies.

From inspection of Fig.~\ref{TempDensity}b, it appears that the dust
temperature does not show any systematic variation with the local
density of either low mass or high mass galaxies.  Further checks to
see whether there is any correlation between either dust temperature
or $\Sigma_5$ with other parameters such as galaxy redshift, physical
size, and luminosity reveal none.

\begin{figure*}
      \includegraphics[width=18cm]{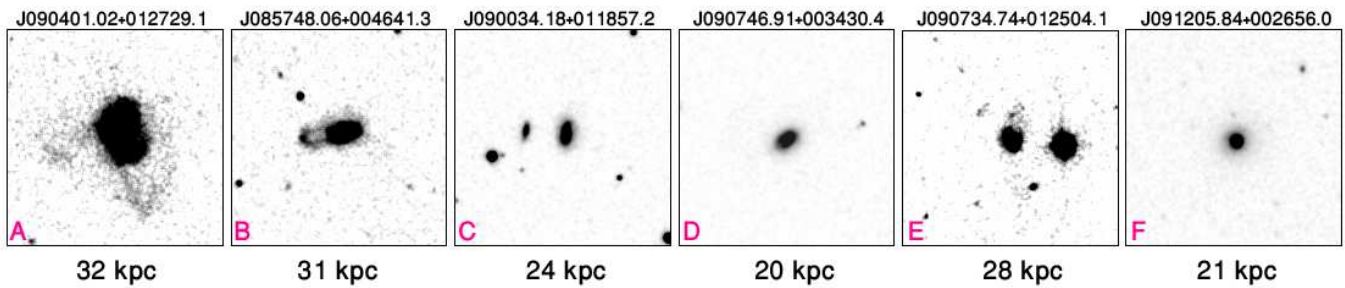}
      \caption{SDSS postage stamp images in the $r$-band 
        ($70\times70$~arcsec$^2$) of galaxies corresponding to the marked
        data points in Fig.~\ref{TempDensity}.  Images are sorted
        (left-to-right) according to the estimated dust temperature.
        The SDSS optical major isophotal diameter in the $r$-band (e.g. {\tt isoA-r}) has been used to measure
 the optical linear diameter of each galaxy $D_{25}$ in units of kpc
 as indicated on each panel.}
          \label{SDSSimages2}
   \end{figure*}

Interestingly, Fig.~\ref{TempDensity} shows the presence of six
outliers (marked as 'A' to 'F') characterized by dust temperatures
($T\approx32\pm2$~K) significantly higher than the values
typically observed in the rest of our sample. SDSS images (see
Fig.~\ref{SDSSimages2}) reveal the presence of diffuse tidal stellar
features associated with at least three (A, C and F) of the six
outliers in Fig.~\ref{TempDensity}. These are the only galaxies in our
sample of 38 objects with obvious signatures of tidal disturbance,
suggesting that the dust temperature has recently been enhanced via
gravitational interaction.

\section{Conclusions}

This study presents a variety of statistics based on {\it Herschel}
observations acquired during the H-ATLAS Science Demonstration Phase.
We started with a sample of SDSS-GAMA galaxies at low redshift
($0.02\lesssim z \lesssim 0.2$) and then characterized the 
properties of the matched submillimeter sources detected in the
H-ATLAS survey from their colours.  Our results can be summarized as follows:
 
\begin{itemize}

\item At least $\approx 50$ per cent of all SDSS galaxies with $r_{_{\rm Petro}} \lesssim 17.0$~mag 
are detectable in H-ATLAS.  In addition, the vast majority
  of detected sources in H-ATLAS are blue/star-forming objects.  The
  majority of undetected faint/blue sources with
  $r_{_{\rm Petro}} > 17.0$~mag remain undetected due to their lower
  signal-to-noise.  By using the NUV$-r$ index as a proxy for galaxy colour, we detect a small fraction ($\sim$ 5 per cent) of red objects (NUV$-r>$~4.5~mag)
  which are mainly galaxies with a higher level of dust attenuation.  As such the
  red sequence population which is dominated by less-dusty/passive objects, do not seem to have a significant
  contribution to the observed submillimeter emission in red objects.
 
 \item The rate of detection of galaxies decreases from $\approx$ 70 per cent
   to 30 per cent as one moves from low density to high density regions.
   This is due to the fact that low density regions are more populated
   by blue/dusty objects. However the detection rate of blue/red
   galaxies remains constant in either low or high density
   environments. In other words the colour of an object rather than its
   local density determines whether it is detectable in H-ATLAS.

 \item The estimated dust temperature in galaxies with detections
   in all SPIRE bands and $\geq3\sigma$ detections in PACS bands is
   $\rm{25\pm4~K}$, regardless of the environment. However, we
   show that gravitationally perturbed systems have temperatures
   significantly higher than the rest of our sample.

 \end{itemize}

The next crucial step is to extend our analysis to a wider range of
galaxy densities/stellar-masses to see how far the environment of
galaxies could affect their observed submillimeter properties. While
{\it Herschel} observations of the Virgo cluster are starting to reveal how
the environment can affect the dust properties of cluster galaxies
\citep{Cortese10}, only a wide-area survey like H-ATLAS will make it
possible to unveil the effects of nurture across the whole range of
densities (from voids to the cores of clusters).  Thus, once completed,
the H-ATLAS survey together with data from the GAMA survey will offer
a unique data set, spanning from the UV to the submm, with which to
investigate the star formation history of galaxies over a large range
of galaxy densities/stellar-masses. At the same time, properties such
as dust temperature and dust-mass/stellar-mass of galaxies can be
explored as a function of environment using PACS/SPIRE submillimeter
fluxes.  This will enable us to further investigate the correlation
between star formation history and dust properties of galaxies as a
function of stellar mass and local density.


\section*{Acknowledgments}

We would like to thank Professor Bahram Mobasher for his constructive comments on our
article.

The {\it Herschel}-ATLAS is a project with {\it Herschel}, which is an ESA space
observatory with science instruments provided by European-led
Principal Investigator consortia and with important participation from
NASA. The H-ATLAS website is http://www.h-atlas.org/.

GAMA is a joint European-Australasian project based around a
spectroscopic campaign using the Anglo-Australian Telescope. The GAMA
input catalogue is based on data taken from the Sloan Digital Sky
Survey and the UKIRT Infrared Deep Sky Survey. Complementary imaging
of the GAMA regions is being obtained by a number of independent
survey programs including {\it GALEX} MIS, VST KIDS, VISTA VIKING, WISE,
{\it Herschel}-ATLAS, GMRT and ASKAP, providing UV to radio coverage. GAMA is
funded by the STFC (UK), the ARC (Australia), the AAO, and the
participating institutions. The GAMA website is:
http://www.gama-survey.org/.

Funding for the SDSS and SDSS-II has been provided by the Alfred
P. Sloan Foundation, the Participating Institutions, the National
Science Foundation, the U.S. Department of Energy, the National
Aeronautics and Space Administration, the Japanese Monbukagakusho, the
Max Planck Society, and the Higher Education Funding Council for
England. The SDSS Web Site is http://www.sdss.org/.

The SDSS is managed by the Astrophysical Research Consortium for the
Participating Institutions. The Participating Institutions are the
American Museum of Natural History, Astrophysical Institute Potsdam,
University of Basel, University of Cambridge, Case Western Reserve
University, University of Chicago, Drexel University, Fermilab, the
Institute for Advanced Study, the Japan Participation Group, Johns
Hopkins University, the Joint Institute for Nuclear Astrophysics, the
Kavli Institute for Particle Astrophysics and Cosmology, the Korean
Scientist Group, the Chinese Academy of Sciences (LAMOST), Los Alamos
National Laboratory, the Max-Planck-Institute for Astronomy (MPIA),
the Max-Planck-Institute for Astrophysics (MPA), New Mexico State
University, Ohio State University, University of Pittsburgh,
University of Portsmouth, Princeton University, the United States
Naval Observatory, and the University of Washington.


\label{lastpage}
\end{document}